\documentclass[preprint]{aastex}
\newcommand       \be           {\begin{equation}}
\newcommand       \ee           {\end{equation}}
\newcommand       \Angstrom     {\,{\rm \AA}}          
\newcommand       \eV           {\,{\rm eV}\,}
\newcommand       \K            {\,{\rm K}}

\newcommand       \cm           {\,{\rm cm}}

\newcommand       \s            {\,{\rm s}}
\newcommand       \erg          {\,{\rm erg}}
\newcommand       \kms		{\,{\rm km \, s}^{-1}}
\newcommand	  \pc		{\,{\rm pc}}

\newcommand       \NH           {N_{\rm H}}

\newcommand       \gtsim        {\gtrsim}
\newcommand       \ltsim        {\lesssim}

\shorttitle{Grains in AGN}
\shortauthors{Weingartner \& Murray}

\begin{document}

\title{X-ray versus Optical Observations of Active
Galactic Nuclei:  Evidence for Large Grains?}

\author{Joseph C. Weingartner \& Norman Murray}
\affil{CITA, 60 St. George Street, University of
Toronto, Toronto, ON M5S 3H8, Canada}
\email{weingart@cita.utoronto.ca, murray@cita.utoronto.ca}

\begin{abstract}

Recently, Maiolino et al.~(2001a, \aap, 365, 28) constructed a sample of 
active galactic nuclei for which both the reddening $E(\bv)$ and the column 
density $\NH$ to the nucleus could be determined.  For most of the 
galaxies in their sample, they found that $E(\bv)/\NH$ is substantially 
smaller than for the diffuse interstellar medium of our Galaxy.  They 
asserted that either the dust-to-gas ratio is lower than in the 
Galaxy or that the grains are so large that they do not extinct or redden 
efficiently in the optical.  
We show that there is no systematic increase in $E(\bv)$ with 
$\NH$ for the Maiolino et al.~(2001a) galaxies, which suggests that the
X-ray absorption and optical extinction occur in distinct media.

Maiolino et al.~(2001b, \aap, 365, 37) suggested that the observed lines of 
sight for the Maiolino et al.~(2001a) galaxies pass through the ``torus'' 
that obscures the broad line region and nuclear continuum in Seyfert 2 
galaxies and argued that the torus grains are larger than Galactic grains.
There is no reason to believe that the lines of sight for these galaxies pass 
through the torus, since the observed column densities are lower than those 
typically observed in Seyfert 2 galaxies.  We suggest instead that the X-ray 
absorption occurs in material located off the torus and/or accretion disk 
while the optical extinction occurs in material located beyond the torus.  
The X-ray absorbing material could either be dust-free or could contain
large grains that do not extinct efficently in the optical.  
There is no conclusive evidence that the grains in 
active galactic nuclei are systematically larger than those in the diffuse  
interstellar medium of our Galaxy.  We discuss an alternative way to probe
the properties of dust in Seyfert tori, but find that observations of 
Seyfert 2 nuclei with higher resolution than currently available will be 
needed in order to place stringent limits on the dust.

\end{abstract}

\keywords{galaxies: active---galaxies: nuclei---galaxies: ISM---dust,
extinction}

\section{Introduction \label{sec:intro}}

The unified model of active galactic nuclei (AGNs) posits that Seyfert 1
(Sy 1) and Seyfert 2 (Sy 2) galaxies are intrinsically the same, but that 
obscuring matter along our line of sight to the nucleus extincts the broad 
lines and nuclear continuum
in Sy 2s (see Antonucci 1993 for a review).  The obscuring matter
is usually referred to as the ``torus'', although its geometry has not 
yet been established.  

The torus is located at a distance $r \ltsim 1 \pc$ from the center, 
on the basis of 
the following observational evidence.  First, the infrared spectral energy 
distribution of AGNs is dominated by thermal emission from grains with a 
range of temperatures extending up to the sublimation temperature, 
$\approx 1500$--$2000 \K$.  This temperature is reached at a distance 
$\ltsim 1 \pc$ from the center.  Variations in the IR continuum 
follow variations in the UV/optical continuum with a delay of 
$\sim$ hundreds of days, confirming the presence of dust with
$r \ltsim 1 \pc$.  See \S 4.3 of Peterson (1997) for further details
and references.  Second, VLBI observations of ``megamasers'' in AGNs 
demonstrate that molecules are located at $r \ltsim 1 \pc$
(see Maloney 2002 for a review).\footnote{The megamaser observations also
suggest a warped disk geometry for the torus; see Maloney (2002).}
Third, Risaliti, Maiolino, \& Salvati (1999) have shown that the column 
density $\NH \gtsim 10^{24} \cm^{-2}$
for at least 50\% of Sy 2 galaxies, and have pointed out that if the 
obscuring matter were located at $r \gtsim 10$--$100 \pc$, then its mass
would exceed the dynamical mass for that region.  

In addition to Sy 1s and Sy 2s, there are also intermediate-type galaxies,
with weaker broad lines than observed in Sy 1s.  For example, Sy 1.8 
galaxies have very weak broad H$\, \alpha$ and H$\, \beta$ lines and 
Sy 1.9s have no detectable broad H$\, \beta$.  Risaliti et al.~(1999)
found that the column density (determined from the photoelectric cutoff in
the X-ray spectrum) along the line of sight to the nucleus
is typically much higher in Sy 2s than in 1.8s and 1.9s (see Figure
\ref{fig:Sy_NH_dist}).  The lower column densities in intermediate-type 
Seyferts could possibly result from our line of sight traversing relatively 
low-density outer regions of the torus.  Alternatively, the X-ray-absorbing
material in Sy 1.8s and 1.9s might be unrelated to the torus.  
Maiolino \& Rieke (1995) found that Sy 1.8s and 1.9s are mostly found in
edge-on host galaxies, implying that the optical extinction is due to dust
located beyond the torus.  If the optical extinction is unrelated to the 
torus, then it seems plausible that the same could be true for the X-ray 
absorption as well.

Maiolino et al.~(2001a; hereafter M01a) presented a sample of 
AGNs for which at least some broad lines are observed, 
but with significant reddening.  For each of the objects in the M01a 
sample, both the reddening $E(\bv)$ and the total H column density $\NH$ 
along the line of sight to the nucleus have been 
estimated.\footnote{The reddening is the extinction at $B$ minus the 
extinction at $V$; i.e., $E(\bv) = A_B - A_V$.} 
The column density is inferred from observations of the absorption of 
X-rays from the central source.
M01a found that, with the exception of a few low-luminosity AGNs, 
$E(\bv)/\NH$ is generally lower than it is in the 
diffuse interstellar medium (ISM) of 
our Galaxy, by factors of a few up to $\sim 100$.  
M01a assumed that the X-ray-absorbing material and optical-absorbing material
are identical, which would imply that either (1) the dust-to-gas ratio in
this material is lower than in the Galaxy or (2) the grains in this material
are so large that, unlike Galactic grains, they do not extinct and redden 
efficiently in the optical.
In a companion paper, Maiolino, Marconi, \& Oliva
(2001b; hereafter M01b) argued in favor of the latter interpretation.
Their argument assumes that the medium probed by M01a is in fact the torus
that is responsible for the optical obscuration of the broad line region 
(BLR) and nuclear continuum in Sy 2s.  

Here, we propose an alternative interpretation, namely, that the lines
of sight to the nuclei in the M01a sample pass not through the torus, but
through lightly or moderately ionized material located off the torus and/or 
accretion disk, perhaps in a wind.  This 
material, which is responsible for the X-ray absorption, 
could be either dust-free or dominated by large enough grains 
that very little extinction and reddening result.  The observed reddening
of the BLR occurs in a dusty medium that is physically distinct from the 
X-ray-absorbing material.  In \S \ref{sec:sample} we describe the M01a 
sample and present evidence that (1) the torus is not being probed and
(2) the X-ray- and optical-absorbing gas are distinct.  
In \S \ref{sec:big_grains} we review the evidence that led M01b to conclude
that the grain size distribution of the obscuring material must be 
dominated by very large grains, and show that this conclusion does not hold
in our alternative scenario.  In \S \ref{sec:extinction_Sy2}, we discuss 
a method for determining the extinction per column in Seyfert tori.  
Currently available observations yield limits that are consistent with 
the torus dust being similar to Galactic dust, but observations at higher 
resolution are needed in order to establish more stringent limits.
We briefly summarize in \S \ref{sec:conclusions}.

\section{The M01a Sample \label{sec:sample}} 

M01a collected a sample of intermediate AGNs for which both the reddening of
the BLR and $\NH$ could be estimated from observations.
They did not attempt to construct an unbiased sample; an object was included
so long as $E(\bv)$ and $\NH$ could both be determined and there was no 
evidence for either ``warm'' absorption (\ion{O}{7} or \ion{O}{8} absorption 
edges) or partial covering of the nucleus.
The reddening was found by comparing the ratio of the fluxes in two different
broad H recombination lines with the expected intrinsic value.  The column 
density was determined from the photoelectric cutoff in the X-ray spectrum.
We reproduce their sample in Table \ref{tab:sample}; see their Table 1 for 
further details.  Note that many of the objects in the M01a sample
are Seyfert 1.8 or 1.9 galaxies.

M01b assumed that the low $E(\bv)/\NH$ found for most of
the objects in the sample is a property of the torus, but this is 
not necessarily true.  The bottom panel in Figure \ref{fig:Sy_NH_dist}
shows the column density distribution for the M01a sample.  The 
distribution mirrors the Risaliti et al.~(1999) distribution for Sy 1.8s
and 1.9s, which may not obscured by the torus, as discussed in 
\S \ref{sec:intro}.   
Of course, column densities characteristic of intermediate-type Seyferts
are to be expected, since some broad lines are observed in the M01a 
sample objects.  

In Figure \ref{fig:EvsNH} we plot $E(\bv)$ versus $\NH$ for the M01a galaxies.
The three points on 
the left side of the figure are the three low-luminosity AGNs for which 
$E(\bv)/\NH$ was found to exceed the Galactic value of $1.7 \times 10^{-22}
\, {\rm mag} \cm^2$ (Bohlin, Savage, \& Drake 1978).  The box is 
AX J0341-44, for which only an upper limit on $E(\bv)$ has been established.

If the gas that absorbs the X-rays were the same as the gas that absorbs 
the optical radiation, then we would expect $E(\bv)$ to increase with 
$\NH$.  However, Figure \ref{fig:EvsNH} is just a scatter plot, suggesting
that the X-ray-absorbing gas and the optical-absorbing gas are, in fact,
unrelated.  The only hint of a correlation between $E(\bv)$ and $\NH$ 
is the presence of three points in the upper right part of the figure.  
However, one of these is AX J0341-44 (with just an upper limit on the 
reddening) and the other two are NGC 1365 and NGC 5506.  Both of these 
latter galaxies are edge-on spirals, and a dust lane is observed to 
directly cover the nucleus in NGC 1365 (see further discussion below).


Closer examination of some of the well-studied galaxies in the M01a 
sample yields valuable insight:


\noindent {\it Mrk 231.}  M01a took their X-ray data on Mrk 231 from 
Turner (1999), who observed it with ROSAT and ASCA.  Maloney \&
Reynolds (2000) have observed it more recently with ASCA, and concluded that
the observed X-rays are reflected and that the direct line of sight to the 
X-ray source is Compton thick ($\NH \gtsim 10^{24} \cm^{-2}$).\footnote{Note 
that \ion{O}{7} and \ion{O}{8} edges would not be observed in this case, 
even if these ions were present, since the X-rays are observed in reflection.}
Gallagher et 
al.~(2002) observed with Chandra and also concluded that the direct line of
sight is Compton thick.  R.~Maiolino (2002, private communication) has
brought our attention to recent (not yet published) XMM-Newton data which 
suggest that Mrk 231 is Compton thin.  We adopt the M01a value for $\NH$
in Figure \ref{fig:EvsNH}, but note that the point representing Mrk 231 
should perhaps be located further to the right.

Gallagher et al.~(2002) also showed that Mrk 231 is a 
broad absorption line quasar (BALQSO), since an HST 
spectrum shows blueshifted \ion{C}{4} absorption.  
Gallagher et al.~interpreted 
their Chandra observations in the context of the radiatively-driven disk wind 
model for BALQSOs (Murray et al.~1995).  They favor a scenario in which the
direct X-rays are absorbed in the highly-ionized ``hitchhiking gas''; the 
observed X-rays are scattered off of the hitchhiking gas and absorbed in 
the BAL wind (see their Figure 12).  If this picture is correct, then the
X-ray-absorbing gas is dust-free, since the BAL wind originates well
within the dust sublimation radius.  We can see this directly from the
observed width $\Delta v$ of the \ion{C}{4} absorption.  The sublimation 
radius $r_{\rm sub} \sim 0.20 L_{46}^{1/2} \pc$, where $L_{46}$ is 
the bolometric luminosity in units of $10^{46} \erg \s^{-1}$ (Laor \& Draine
1993).  If the absorption occurs at $r_{\rm sub}$, then we expect the 
width to be $\Delta v_{\rm sub} \sim (2 G M_{\rm BH}/r_{\rm sub})^{1/2}$,
where $M_{\rm BH}$ is the mass of the black hole.  Thus, 
\be
\Delta v_{\rm sub} \sim 2.1 \times 10^3 M_8^{1/2} L_{46}^{-1/4} \kms~~~,
\ee
where $M_8$ is $M_{\rm BH}$ in units of $10^8$ solar masses.  
The infrared luminosity of Mrk 231 is $\sim 1.5 \times 10^{46} \erg \s^{-1}$
(Rieke \& Low 1972, 1975)
and, from the Gallagher et al.~spectrum, $\Delta v \gtsim 9 \times 10^3 
\kms$.  Thus, $\Delta v > \Delta v_{\rm sub}$, and the BAL wind is within
the sublimation radius.  

Mrk 231 is also an ultraluminous 
infrared starburst galaxy and contains $\approx 5$ times as much molecular
gas as the Galaxy (Sanders et al.~1987).  Thus, a reddening $E(\bv) \approx
0.34$ due to dust in the host galaxy would not be unexpected.

\noindent {\it NGC 1365.}  Risaliti,
Maiolino, \& Bassani (2000) observed NGC 1365 with BeppoSAX.  Combining their 
observation with a previous ASCA observation (Iyomoto et al.~1997), 
Risaliti et al.~concluded that there must be a directly observed component
and a reflected component.  They found that 
$\NH \approx 4 \times 10^{23} \cm^{-2}$ for the directly transmitted X-rays
and that a Compton-thick component is required near the central source to 
account for the high reflection efficiency.  NGC 1365 is an edge-on spiral 
galaxy; Edmunds \& Pagel (1982) suggested that most of the BLR extinction 
occurs in a dust lane that has been observed to just barely extend over the 
nucleus.

\noindent {\it IRAS 13197-1627 (= MCG-03-34-064).}  M01a took broad 
H$\, \alpha$ and H$\, \beta$ fluxes from Ag\"{u}ero et al.~(1994).  
However, other authors have reported only narrow lines.  Young et 
al.~(1996) concluded that the narrow lines have broad wings and did not
find a need for separate broad lines.  Thus, this object (with 
$E(\bv)/\NH \approx 0.0092$ relative to Galactic, the lowest of the 
M01a AGNs) should be removed from the M01a sample.

\noindent {\it NGC 5506.}  M01a obtained their estimate for the reddening
from the ratio of near-infrared lines, taking Pa$\, \beta$ from Rix 
et al.~(1990).  However, Goodrich, Veilleux, \& Hill (1994) argue that there 
is actually
no broad component of the Pa$\, \beta$ line, just broad wings on the narrow 
component.  Young et al.~(1996) found that the narrow lines are broader in
polarized light, supporting the view of Goodrich et al.  From the absence 
of broad Pa$\, \beta$, Goodrich et al.~conclude that $E(\bv) \gtsim 3.5$,
a factor of 2.2 greater than the reddening adopted by M01a.  Given that
M01a already found $E(\bv)/\NH = 0.27$ Galactic, the discrepancy between
NGC 5506 and the Galaxy is small, if there is any discrepancy at all. 
Young et 
al.~argue that NGC 5506 is actually a Sy 1 that is heavily reddened by 
dust in the host galaxy, which is of type S0/a and is seen edge-on
(see p.~1242 of their paper for the details of their evidence).  
 
\noindent {\it NGC 2992.}  M01a take the broad line reddening from 
Gilli et al.~(2000).  Gilli et al.~find the same reddening for the 
narrow line region (NLR) as
for the BLR, and suggest that both are reddened by the same medium, 
perhaps the dust lane that is observed to run across the nucleus.

\noindent {\it MCG-5-23-16.}  M01a adopt $E(\bv) = 0.6$ from the broad
Pa$\, \alpha$ and Br$\, \gamma$ fluxes measured by Veilleux, Goodrich, 
\& Hill (1997).  Veilleux et al.~note that it is not clear whether or not
broad components of the Balmer lines have been detected.  Taking a 
generous upper limit on the broad H$\, \alpha$ flux, they find upper limits
on the BLR $E(\bv)$ of 2.3 (1.9) from Pa$\, \beta$/H$\, \alpha$
(Br$\, \gamma$/H$\, \alpha$).  They do not understand the discrepancy with
the result derived using only the infrared lines.  For the NLR, 
Durret \& Bergeron (1988) found $E(\bv)=0.85$ from 
H$\, \alpha$/H$\, \beta$ and Goodrich et al.~(1994) found $E(\bv)=0.91$
(0.77) from H$\, \alpha$/H$\, \beta$ (Pa$\, \beta$/H$\, \alpha$).
Ferruit, Wilson, \& Mulchaey (2000) imaged MCG-5-23-16 with HST and found
a rich dust structure in the vicinity of the nucleus, on the scale of 
hundreds of pc.  They also resolve the NLR, which extends to 
$\approx 180 \pc$ from the nucleus in two opposite directions.  Based on color 
variations across their images, they tentatively find $E(\bv) \approx 0.8$
for the nucleus, consistent with the above determinations for the NLR.
If the BLR $E(\bv)=0.6$, then it appears that the reddening is due to dust
located in the host galaxy, well beyond the torus.  If the BLR 
$E(\bv) \gtsim 2$, then this may still be the case, but it could also be
that some of the reddening arises in material much closer to the nucleus.

\noindent {\it IRAS 05189-2524.}  M01a take the reddening from 
Severgnini et al.~(2001), who found $E(\bv)=0.7$ for the BLR from 
Pa$\, \alpha$/Pa$\, \beta$.  Severgnini et al.~also report the fluxes in
the narrow lines, and their ratio yields a NLR reddening that is a factor
of 2.8 times {\it larger} than the BLR reddening.  If this is correct,
then it certainly suggests that dust in the host galaxy is responsible.
However, Veilleux, Sanders, \& Kim (1999), who also measured broad 
Pa$\, \alpha$ and Pa$\, \beta$, caution against using Pa$\, \beta$ to 
determine the reddening, because its profile is affected by strong 
absorption features arising in Earth's atmosphere.  Veilleux et al.~(1999)
find $E(\bv) \gtsim 2.8$ for the BLR from Pa$\, \alpha$ and the upper limit 
on H$\, \alpha$ from Veilleux et al.~(1997).  Veilleux et al.~(1999) find
$E(\bv) = 2.6$ (2.0) for the NLR from Pa$\, \beta$/H$\, \alpha$
(H$\, \alpha$/H$\, \beta$).  

\noindent {\it NGC 526a.}  Winkler (1992)
found that the NLR is unreddened and did not detect broad lines.  Winkler
mentions that an unpublished spectrum taken by Martin Ward in 1978 did 
detect broad H$\, \alpha$ and H$\, \beta$, and that the line ratio 
indicated substantial reddening.  Thus, this object might be a case in 
which the BLR-obscuring dust is located in the vicinity of the torus.
M01a cite Marconi et al.~(in preparation) for the reddening; this paper
apparently has not yet been published.

To summarize this section, a plot of $E(\bv)$ versus $\NH$ for the M01a 
sample suggests that the X-ray absorption and optical extinction arise in 
different material.  Observations of several of the M01a galaxies reveal
that the optical extinction arises at $r \gtsim 100 \pc$, beyond the 
location of the torus.  This is consistent with the following two results: 
(1) most Sy 1.8s and 1.9s are located in edge-on galaxies (Maiolino \& Rieke 
1995) and (2) the nuclear reddening in Sy 1s increases with the inclination
angle of the host galaxy (Crenshaw \& Kraemer 2001 and references therein).
In a couple of cases, there are also indications
that the X-ray-absorbing gas is located close to the center, at 
$r \ltsim 1 \pc$.  Finally, the low inferred column densities suggest that
the line of sight does not pass through the torus, or at most only through
a relatively low density outer layer of the torus.  Taken together, these
pieces of evidence strongly contradict the M01a and M01b interpretation 
that $E(\bv)/\NH$ is substantially lower than Galactic for AGN tori.  
Rather, the evidence suggests that the M01a AGNs suffer optical extinction
farther out in the host galaxy (on scales of $\sim 0.1$--$1 \, {\rm kpc}$)
and X-ray absorption in close to the nucleus, in material that lacks 
typical Galactic grains, and that the obscuring torus is not being probed 
in these objects.

\section{Large Grains? \label{sec:big_grains}}

M01b considered various possible explanations for the low 
$E(\bv)/\NH$ found in the M01a sample and dismissed most of them.  
For example, they considered the possibility that the broad lines are not
viewed directly through the same material that absorbs the X-rays, but
rather are seen in reflection from a medium located at some distance from 
the center, with a lower column of material between us and the scattering
medium.  However, the observed polarization of the broad lines is 
usually much lower than would be expected in this case.  Furthermore,
the generally low scattering efficiency would imply unexpectedly large 
intrinsic broad line intensities, given the observed fluxes.

M01b also considered the possibility that the line of sight to the 
central X-ray source encounters different material than the line of sight
to the BLR.  Although this could certainly occur, it is unlikely to be
the case for all of the galaxies discussed by M01a, since the distance 
from the center to the BLR is much less than the distance to the torus.

Perhaps the most obvious explanation is that the dust-to-gas ratio in the
obscuring material is lower than it is in the Galaxy.  M01b argue that 
this cannot be the case for all of the M01a galaxies as follows.
As mentioned above, they assume that the 
obscuring medium is the torus, which presumably consists largely of neutral
gas.  Thus, most of the H-ionizing ultraviolet photons will be absorbed,
either by dust or by gas.  In the latter case, emission in H recombination
lines results.  If the torus has a covering factor of 
$\sim 80\%$ (see below for further discussion of this), then in the 
dust-depleted case a large fraction of the ultraviolet continuum will be 
reprocessed into H recombination lines, with widths comparable to the 
narrow line widths.  On the other hand, the narrow line intensities 
imply a NLR covering factor of $\sim 8\%$.  Thus, the dust depletion
scenario would lead to $\sim 10$ times too much narrow line emission.

M01b argued that this problem can be avoided if the dust is dominated by 
large grains.  The grains are large enough that the extinction efficiency 
is low in the optical, yielding the low $E(\bv)/\NH$, yet are still able
to absorb enough of the UV continuum to suppress the H recombination lines.

As discussed in \S \ref{sec:sample}, the torus most likely
does not lie along our line of sight to the nucleus for the M01a galaxies.
We agree with M01b that ionizing photons incident on the torus are absorbed 
by grains, but in our scenario these grains need not be different
from Galactic grains.  
We can estimate the covering factor of the X-ray-absorbing material probed  
by the M01a sample from the fraction of Seyferts that are of 
intermediate type.  Maiolino \& Rieke (1995) found that the ratio
Sy 2 : Sy (1.8+1.9) : Sy (1+1.2+1.5) $\approx$ 3:1:1 and concluded that
the intermediate Seyferts usually are obscured not by the torus proper, but 
by material in the host galaxy located at a greater distance from the nucleus.
This led M01b to take the covering factor of neutral material to be 80\%. 
We conclude that the covering factor of the X-ray-absorbing material probed 
by the M01a sample is as low as 20\% (if it does not overlap the torus)
and as high as 80\% (if it also covers the 60\% of the sky covered by the 
torus).  In any case, the effective covering factor is 20\%, since any line
emission originating from behind the torus would not be visible.
Thus, if this material were neutral and dust-free, then the excess H 
recombination line intensity is reduced from a factor of $\sim 10$
to a factor of $\sim 2.5$.  

However, the H in this material is probably ionized.  For dust-free gas,
the largest possible ionized column density can be found by balancing 
ionization and recombination:
\be
\NH \approx 6 \times 10^{23} \cm^{-2} \left( \frac{\dot{N}}{10^{54} \s^{-1}}
\frac{10^{-13} \cm^3 \s^{-1}}{\alpha_r} \frac{n}{10^5 \cm^{-3}}
\right)^{1/3}~~~,
\ee
where $\dot{N}$ is the rate at which H-ionizing photons are emitted from 
the central source and $\alpha_r$ is the case B radiative recombination 
coefficient.  We estimate $\dot{N} \sim 10^{54} \s^{-1}$ by taking an 
ionizing luminosity of $10^{44} \erg \s^{-1}$ and an average photon energy
of $50 \eV$.  For $T=10^4 \K$ ($10^5 \K$), $\alpha_r = 4.2 \times 10^{-13}$
($6.9 \times 10^{-14} \cm^3 \s^{-1}$).\footnote{We evaluate $\alpha_r$ using
the FORTRAN routine {\it rrfit}, written by D.~A.~Verner, and available 
at http://www.pa.uky.edu/$\sim$verner/fortran.html.}  Thus, we expect that
the central source can typically maintain an ionized column density of 
$\NH \approx 6 \times 10^{23} \cm^{-2}$, which is a factor $\approx 2$ 
larger than the highest $\NH$ in the M01a sample (for IRAS 13197-1627, 
which, as we showed in \S \ref{sec:sample}, should be removed from the M01a 
sample, or NGC 1365).  In other words, for NGC 1365, only half of the 
ionizing photons are absorbed by the gas, assuming it is dust-free.
The other half could either be absorbed by dust farther out in the host 
galaxy or escape the host galaxy altogether.  Since significant reddening
is observed for the M01a galaxies, the former alternative is generally 
more likely.  In either case, only half of the ionizing photons passing 
through this medium result in H recombination lines.  Thus, for NGC 1365,  
the excess narrow line power is reduced from a factor 2.5 to a 
factor 1.25.  For the other galaxies in the M01a sample, with lower $\NH$,
a smaller
(often much smaller) fraction of the ionizing photons are absorbed, and 
there is no excess narrow line power.  Thus, in our scenario there is no 
need for larger-than-Galactic grains in active galactic nuclei.  They are 
not needed in the torus, since the M01a observations do not probe the 
torus and hence do not rule out Galactic-type grains.  They are not needed
in the material probed by the M01a sample, because this material is ionized 
and has a lower covering factor than the torus.
   
M01a pointed out two other observations that they claim argue against the 
presence 
of typical Galactic grains in the obscuring media in AGNs.  First, the 
$2175 \Angstrom$ extinction feature, which is nearly ubiquitous in the 
Galaxy and is usually attributed to carbonaceous grains, has not been 
observed in those Sy 1s for which some extinction has been measured.   
Li \& Draine (2001) and Weingartner \& Draine (2001)
suggested that the $2175 \Angstrom$ feature is due to polycyclic aromatic 
hydrocarbon (PAH) molecules.  PAHs can be destroyed in the harsh radiation
environment of AGNs (see, e.g., Voit 1992), so the absence of the 
$2175 \Angstrom$ feature in Sy 1s is perhaps not surprising.  It is interesting
to note that the $2175 \Angstrom$ feature is also absent in the SMC bar 
(Gordon \& Clayton 1998).  In intermediate and type 2 Seyferts, the 
extinction is so strong that the UV emission is completely suppressed,
and there is no chance to search for the $2175 \Angstrom$ feature.  
Since the extinction in Sy 1s is not due to the torus, the 
lack of the $2175 \Angstrom$ feature in these objects does not tell us about 
the dust in the torus.  

M01a also noted that the $9.7 \micron$ feature (associated with silicate
grains with sizes $\ltsim 3 \micron$) is generally very weak in AGNs.  
Clavel et al.~(2000) found that this feature appears weakly in emission 
in Sy 1s and may appear in either emission or absorption in Sy 2s, in 
either case very weakly.  It is difficult to determine the silicate feature
accurately because of strong PAH emission features on either side of it.
Clavel et al.~inferred that there is substantial mid-infrared extinction
in Sy 2s; M01a concluded that there must therefore be grains in the torus,
but that they must be large enough to not produce a $9.7 \micron$ absorption
feature.  We consider this to be the best piece of evidence for large grains
in Seyfert tori.  Unlike the M01a sample, the Clavel et al.~observations
apparently do probe the torus.  M01a noted that radiative transfer effects
could possibly lead to the weak silicate feature in Sy 2s, but considered
this unlikely.  In this context, it is interesting to note that earlier 
observations (that did not take into account the PAH absorption near 
$9.7 \micron$) found the $9.7 \micron$ feature to be absent in Sy 1s but 
strongly present in absorption in Sy 2s (Roche et al.~1991).  Although 
this was initially perplexing, Manske et al.~(1998) were able to find a 
simple radiative transfer scheme that could account for the observations.
Further study is needed to determine whether the Clavel et 
al.~observations can be explained by a radiative transfer scheme when 
silicate grains with $a \ltsim 3 \micron$ are present.  

\section{Extinction in Sy 2s \label{sec:extinction_Sy2}}

As discussed above, the study of reddening in intermediate-type Seyferts 
probably
does not inform us about the properties of grains in the obscuring torus.
Perhaps an analysis of extinction in Sy 2s could be helpful.  
Narrow-aperture observations of the nucleus (e.g., in the optical or 
near-infrared) would yield fluxes which can be compared with the flux  
expected in the case of no obscuration, using an isotropic indicator 
of the intrinsic luminosity (e.g., [\ion{O}{3}] $\lambda 5007$; see Bassani 
et al.~1999).  This would yield a lower limit on the extinction due to the
torus, since some of the observed flux could be due to scattering of the
nuclear radiation and/or to emission by material located beyond the torus.  

Quillen et al.~(2001) compiled high-resolution HST images of Seyfert galaxies
at $1.6 \micron$ and tabulated the nuclear fluxes at this wavelength.  
In this section, we use their data to try to constrain $A_{1.6}/\NH$
for Seyfert tori ($A_{1.6}$ is the extinction at $1.6 \micron$).  Note
that $1.6 \micron$ is approximately the shortest wavelength at which 
emission from hot dust can be observed; if the dust were hot enough to 
emit at shorter wavelengths, then it would have sublimated away.
Thus, observations at somewhat shorter wavelengths, if available, would
be preferable.  

In Table \ref{tab:Sy1} we list $1.6 \micron$ fluxes for all of the Sy 
1+1.2+1.5 galaxies from Quillen et al.~for which [\ion{O}{3}] fluxes are 
available from Whittle (1992).  In Figure \ref{fig:L1.6vsOIII} we plot
the $1.6 \micron$ luminosities versus the [\ion{O}{3}] luminosities for 
these galaxies (squares; cf Figure 3a in Quillen et al.).  
We infer the distances to the galaxies from 
the radial velocities, assuming the Hubble constant $H = 60 \kms \,
{\rm Mpc}^{-1}$.  A least squares fit yields the expected intrinsic 
$1.6 \micron$ luminosity $L_{1.6}^{\rm int}$ as a function of the observed
[\ion{O}{3}] luminosity $L_{\rm O}$:  
$L_{1.6}^{\rm int} = 2.83 \times 10^{11} 
L_{\rm O}^{0.766}$ (with the luminosities in $\erg \s^{-1}$).  
Note the large scatter of the observations about the fit, due in part to the 
large uncertainties in the [\ion{O}{3}] fluxes (Whittle 1992).  

In Table \ref{tab:Sy2} we list the Sy 2 galaxies from Quillen et al.~for
which [\ion{O}{3}] fluxes and $\NH$ (measurements or lower limits) are 
available from Bassani et al.~(1999) or Risaliti et al.~(1999).  We also
include these Sy 2s in Figure \ref{fig:L1.6vsOIII}; note that they mostly 
lie well below the Sy 1s.  We estimate $A_{1.6}$ by comparing the observed
$1.6 \micron$ luminosity, $L_{1.6}^{\rm obs}$, with $L_{1.6}^{\rm int}$ and 
compare 
$A_{1.6}/\NH$ with the typical value for the diffuse ISM in our Galaxy, 
$1.0 \times 10^{-22} \cm^2$ (Cardelli, Clayton, \& Mathis 1989); see
Table \ref{tab:Sy2}.  In cases where only an upper limit on the $1.6 \micron$
flux was established, a lower limit on $A_{1.6}/\NH$ results.  Similarly,
upper limits on $A_{1.6}/\NH$ are established when only lower limits on 
$\NH$ are available.  

The resulting $A_{1.6}/\NH$ (normalized to the typical Galactic value)
cover a range similar to the Galactic-normalized $E(\bv)/\NH$ found by M01a 
for intermediate Seyferts.  We suspect that the large values obtained for
NGC 1672 and NGC 3079 may be due to underestimates of $\NH$.  In Figure
\ref{fig:A1.6vsNH}, we plot $A_{1.6}$ versus $\NH$.  As for the M01a 
sample, $A_{1.6}$ does not systematically increase with $\NH$, indicating
that we have not actually determined $A_{1.6}/\NH$ for the torus.  For 
most of the objects, $L_{1.6}^{\rm obs} \ltsim 0.1 L_{1.6}^{\rm int}$
(i.e., $A_{1.6} \gtsim 2.5$).  Since the HST resolution at the distances 
of the observed galaxies is $\gtsim 10 \pc$, scattering at locations 
beyond the torus could be responsible for most of the observed $1.6 \micron$
flux.  In other words, the determinations of $A_{1.6}/\NH$ in this section
should actually be interpreted as lower limits.
The large scatter of the Sy 1 galaxies about the 
$L_{1.6}$--$L_{\rm O}$ relation indicates that inferred values of 
$A_{1.6} \ltsim 2$ are suspect.  Also, galaxies with low inferred 
$A_{1.6}$ might have an unresolved star cluster very close to the nucleus 
(Quillen et al.~2001).  Thus, the results in this section are entirely 
consistent with the dust in Seyfert tori being similar to Galactic dust.
More stringent limits on the extinction per column will require higher 
resolution.  Improved measurements of [\ion{O}{3}] fluxes (for both Sy 1s 
and 2s) would also be helpful.

\section{Summary \label{sec:conclusions}}

M01a constructed a sample of intermediate-type Seyfert galaxies 
that have been observed in the X-ray and in the optical or infrared and 
found that $E(\bv)/\NH$ is substantially lower than its value in the Galaxy 
for most of the objects in the sample.  M01b concluded that the only viable
explanation for this result is that the grains in Seyfert tori are 
typically substantially larger than the grains in the ISM of our Galaxy.

\begin{figure}
\epsscale{1.00}
\plotone{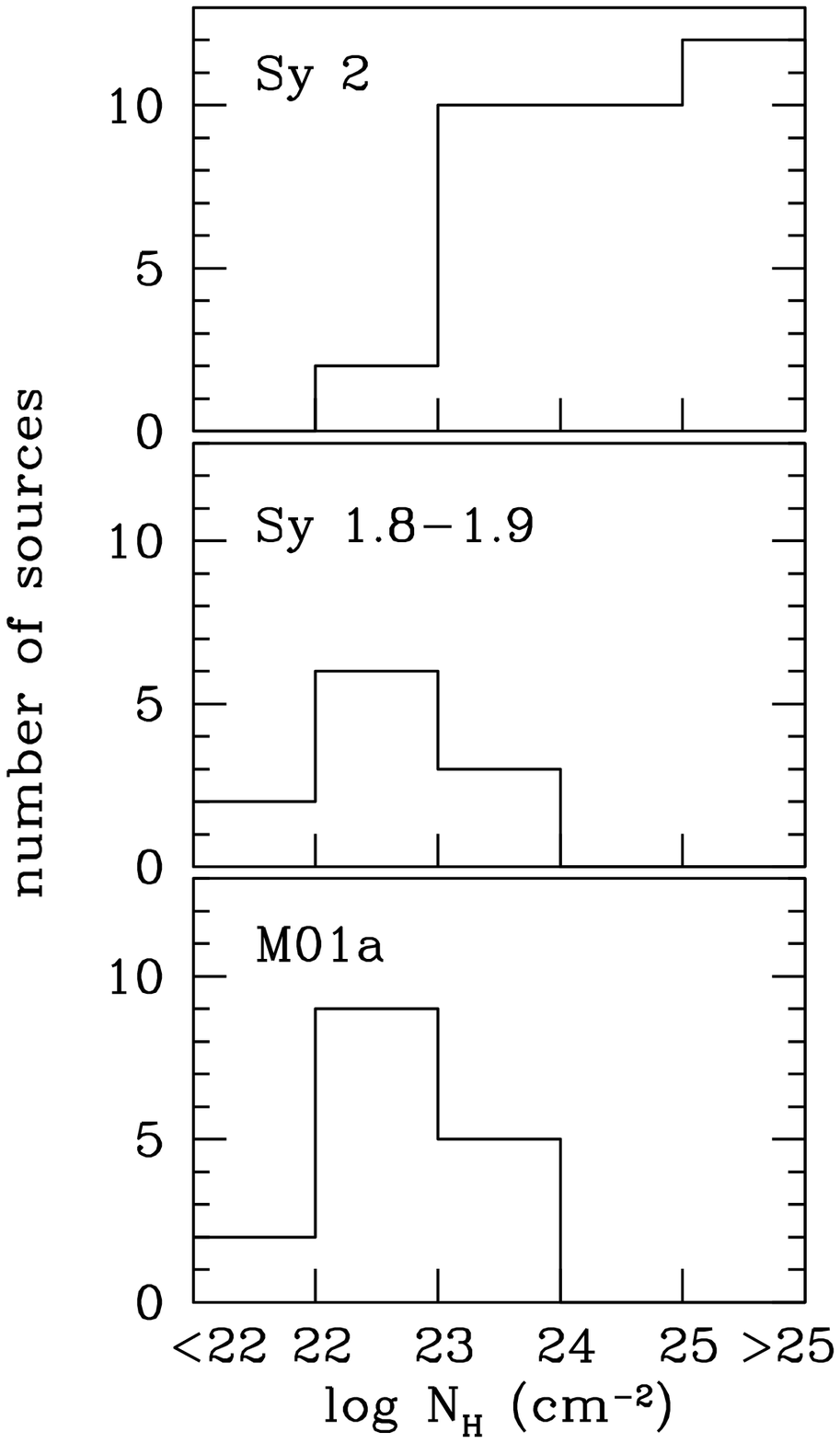}
\caption{
\label{fig:Sy_NH_dist}
X-ray-absorbing column density ($\NH$) distribution for Sy 2 and 
intermediate Sy (1.8--1.9) galaxies in the ``optimal'' sample of Risaliti 
et al.~(1999) and for the galaxies in the M01a sample.
        }
\end{figure}

\begin{figure}
\epsscale{1.00}
\plotone{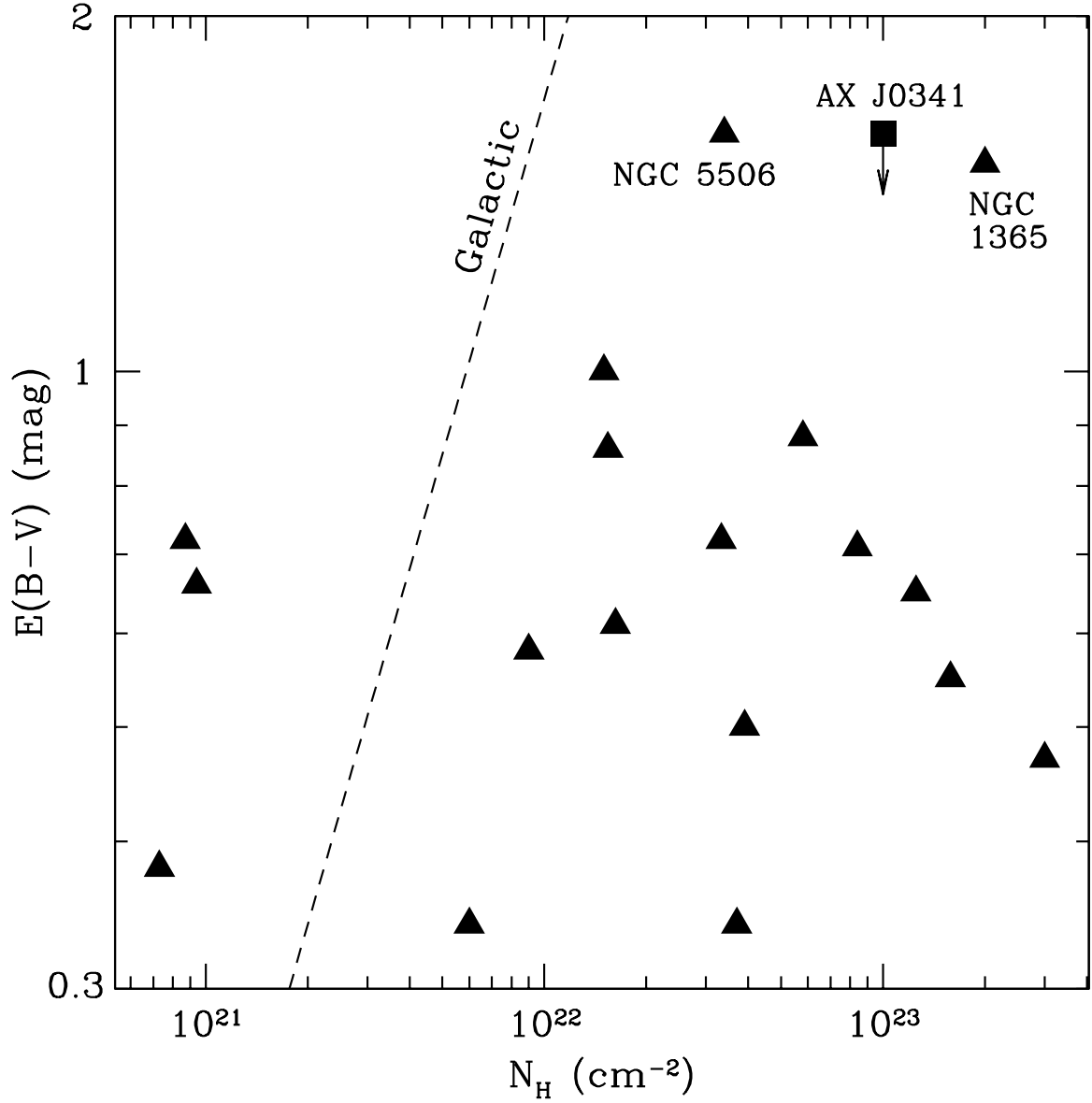}
\caption{
\label{fig:EvsNH}
Reddening versus column density for the objects in the M01a sample.  The 
box corresponds to AX J0341-44, for which only an upper limit has been 
established for the reddening.  The dashed line is the average reddening 
for the diffuse ISM in our Galaxy.
        }
\end{figure}

\begin{figure}
\epsscale{1.00}
\plotone{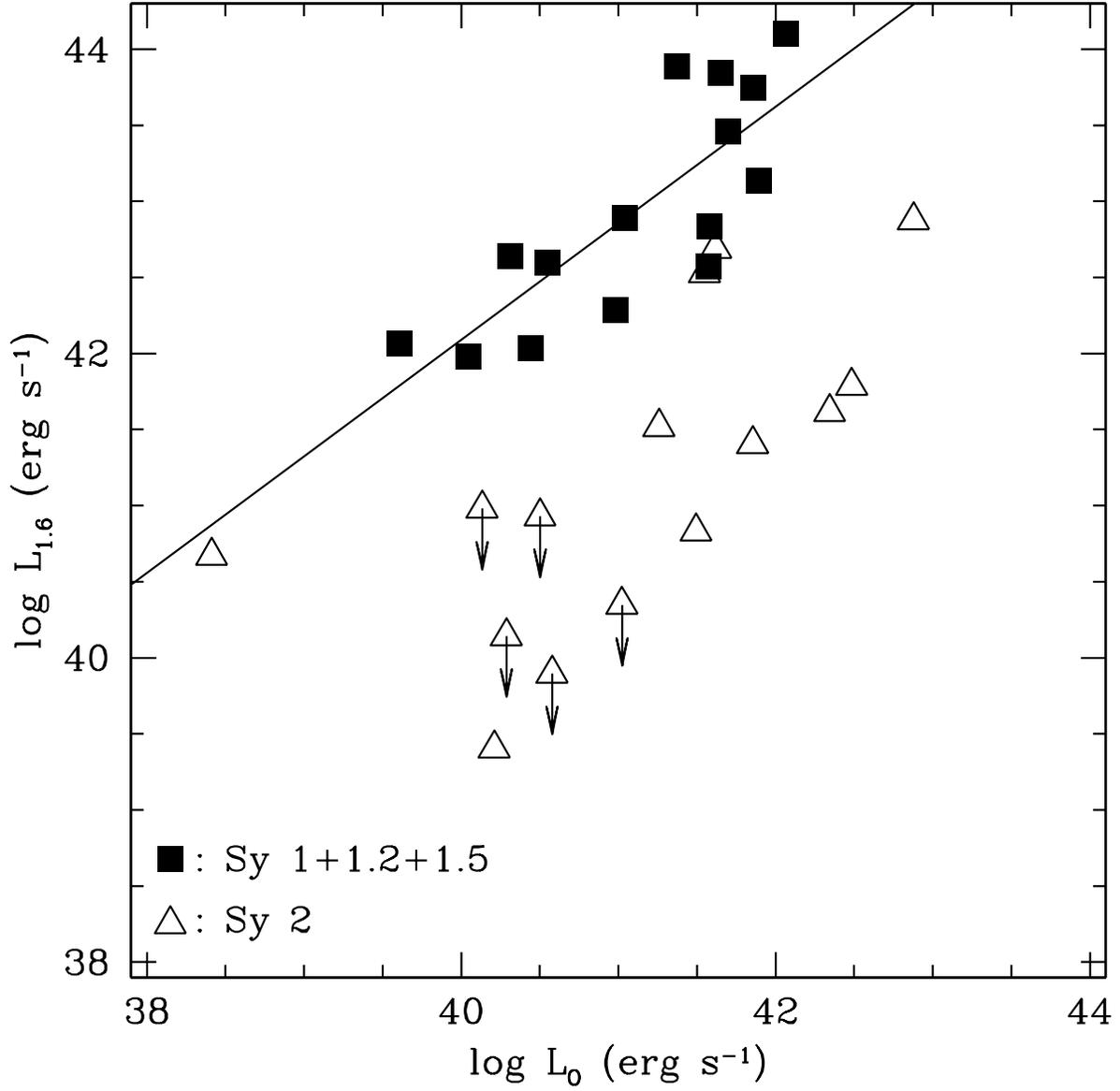}
\caption{
\label{fig:L1.6vsOIII}
Luminosity at $1.6 \micron$ versus [\ion{O}{3}] $\lambda 5007$ luminosity 
for Seyfert galaxies in Tables \ref{tab:Sy1} and \ref{tab:Sy2}.  The solid
line is a least squares fit for Sy 1+1.2+1.5 galaxies.  Arrows indicate
upper limits.
        }
\end{figure}

\begin{figure}
\epsscale{1.00}
\plotone{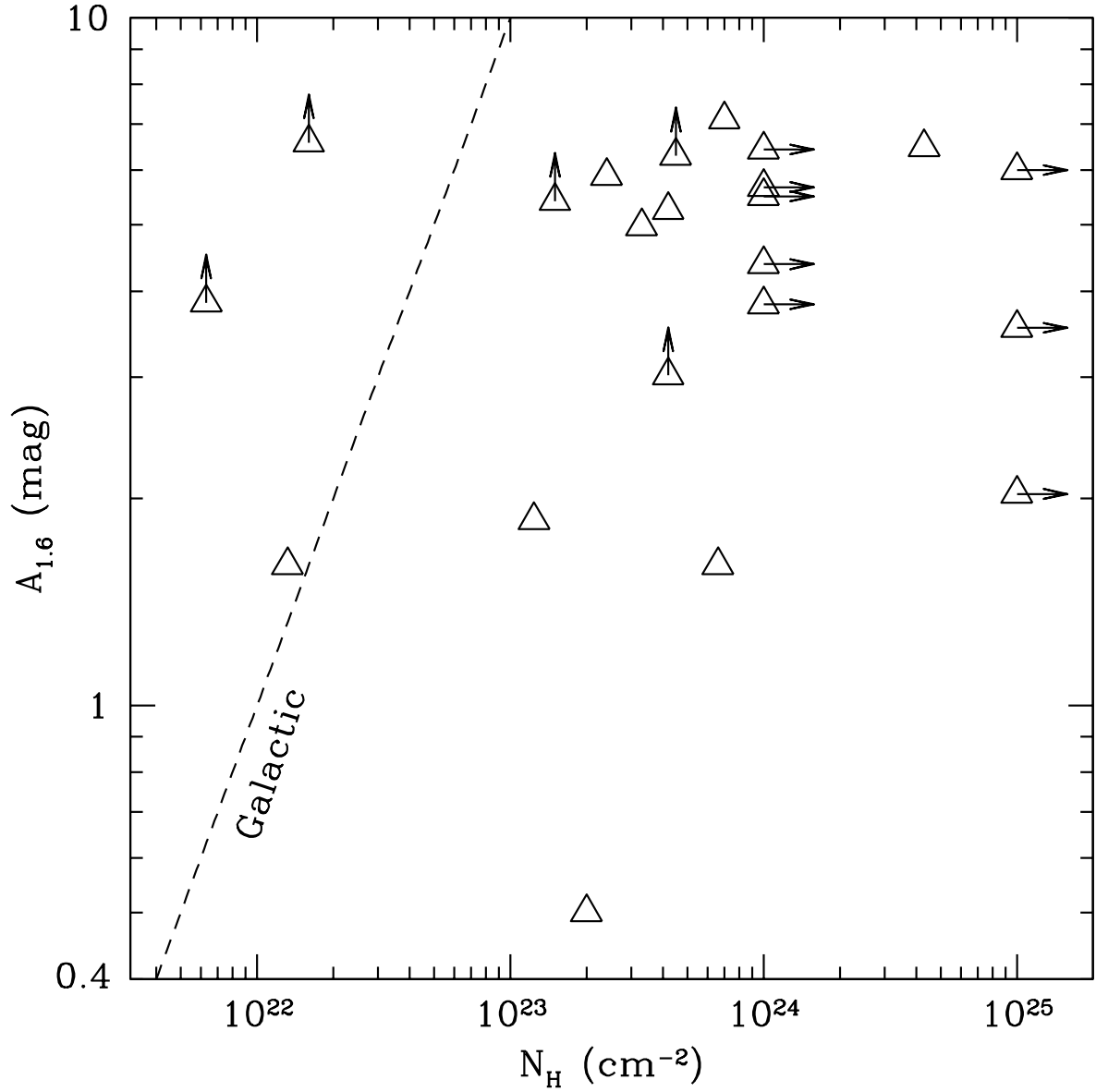}
\caption{
\label{fig:A1.6vsNH}
Inferred extinction at $1.6 \micron$ (see \S \ref{sec:extinction_Sy2})
versus column density for Sy 2s from Table \ref{tab:Sy2}.  Arrows 
indicate lower limits.  The dashed curve is the $1.6 \micron$ extinction
for the diffuse ISM in our Galaxy. 
        }
\end{figure}

Here, we have shown that the material that absorbs the X-rays is probably 
unrelated to the material that absorbs the optical/infrared radiation
(\S \ref{sec:sample}, Figure \ref{fig:EvsNH})
and that the torus probably is not probed by the observations of the 
M01a sample (\S \ref{sec:sample}, Figure \ref{fig:Sy_NH_dist}).
We suggest that, alternatively, the line of sight towards 
an M01a galaxy passes through ionized material located just off the 
torus and/or the accretion disk.  This material is responsible for the 
X-ray absorption, while the optical/infrared extinction occurs in material 
farther from the nucleus, where the dust may be quite similar to Galactic
dust.  The X-ray-absorbing material may be dust-free or may contain large
grains that have very small extinction efficiencies in the optical/infrared
(\S \ref{sec:big_grains}).
This material may be associated with a disk wind, which would originate 
within the dust sublimation radius.  In this case, the material would 
naturally be dust-free.  The disk wind scenario can be tested by searching for 
blue-shifted absorption by ions characteristic of lightly or moderately 
ionized gas, e.g., \ion{C}{4}.  However, such ultraviolet absorption lines 
could be difficult to detect due to the moderate levels of extinction.

Although the M01a sample does not present evidence in favor of large grains
in AGN tori, nor does it present evidence against large grains.  
The apparent lack
of strong $9.7 \micron$ silicate absorption in Sy 2s could indicate large
grains in Seyfert tori, as pointed out by M01a.  More detailed studies of
radiative transfer are needed in order to clarify this point.

We have attempted to obtain limits on the $1.6 \micron$ extinction per column 
in Sy 2s
(\S \ref{sec:extinction_Sy2}), but the HST does not have adequate 
resolution to exclude nuclear light that is scattered into our line of 
sight from locations beyond the torus.  Current observations are consistent 
with the torus dust being similar to Galactic dust; observations at 
higher resolutions will yield more stringent limits.

\acknowledgements
We are grateful to R.~W.~Goodrich, R.~Maiolino, and A.~C.~Quillen for 
helpful discussions. 
This research was supported by NSERC of Canada 
and has made use of the SIMBAD database, operated at CDS, 
Strasbourg, France and the NASA/IPAC Extragalactic Database (NED) which is 
operated by the Jet Propulsion Laboratory, California Institute of 
Technology, under contract with the National Aeronautics and Space 
Administration.


\begin{deluxetable}{ccccc}
\tablewidth{0pc}
\tablecaption{M01a Sample of AGNs
\label{tab:sample}}
\tablehead{
\colhead{Name}&
\colhead{$E(\bv)$}&
\colhead{$\NH$}&
\colhead{$E(\bv)/\NH$}&
\colhead{Type\tablenotemark{a}}
\\
\colhead{}&
\colhead{mag}&
\colhead{$10^{20} \cm^{-2}$}&
\colhead{rel.~Gal.\tablenotemark{b}}&
\colhead{}
}
\startdata
M 81			&0.66	 &9.4	&4.11		&1.8\\
NGC 4639		&0.38 	 &7.3	&3.04		&1.8\\	
NGC 5033		&0.72	 &8.7	&4.84		&1.9\\
NGC 1365		&1.5	 &2000	&0.022		&1.8\\
Mrk 231			&0.34	 &370	&0.054		&1\\
IRAS 13197-1627		&0.47	 &3000	&0.0092		&1.8\\
SAX J1519+65		&0.55	 &1580	&0.020		&1.9\\
NGC 5506		&1.59	 &340	&0.27		&1.9\\
NGC 2992		&0.58	 &90	&0.38		&1.9\\
SAX J0045-25		&0.5	 &390	&0.075		&1.9\\
Mrk 6			&0.72	 &333	&0.13		&1.5\\
MCG-5-23-16		&0.61	 &162	&0.22		&2\\
SAX J1218+29		&0.65	 &1250	&0.030		&1.9\\
IRAS 05189-2524		&0.71	 &840	&0.049		&2\\
NGC 526a		&1.0	 &150	&0.39		&1.9\\
3C 445			&0.88	 &580	&0.088		&1\\
SAX J1353+18		&0.86	 &154	&0.33		&\\
AX J0341-44		&$<1.59$ &1000	&$<0.09$	&\\
PG 2251+11		&0.34	 &60	&0.33		&\\
\enddata
\tablenotetext{a}{Seyfert type, if available.  From the NASA/IPAC 
Extragalactic Database (NED), with the following exceptions:
SAX J1519+65, J0045-25, and J1218+29 (Maiolino et al.~2000); 
NGC 2992 (Gilli et al.~2000); NGC 526a (Landi et al.~2001).}
\tablenotetext{b}{Observed values of $E(\bv)/\NH$ are given relative to
the Galactic value of $E(\bv)/\NH = 1.7 \times 10^{-22} \, {\rm mag} \cm^2$.
For comparison, the mean Galactic column density, for given latitude 
$|b| > 2\arcdeg.5$, is $\NH = 2.9 \times 10^{20} \csc |b| \cm^{-2}$
(Dickey \& Lockman 1990).}
\end{deluxetable}

\begin{deluxetable}{cccc}
\tablewidth{0pc}
\tablecaption{Seyfert 1+1.2+1.5 Galaxies
\label{tab:Sy1}}
\tablehead{
\colhead{Name}&
\colhead{[\ion{O}{3}] flux\tablenotemark{a}}&
\colhead{$1.6 \micron$ flux\tablenotemark{b}}&
\colhead{$v_{\rm r}$\tablenotemark{c}}
\\
\colhead{}&
\colhead{$10^{-11} \erg \s^{-1} \cm^{-2}$}&
\colhead{mJy}&
\colhead{$\kms$}
}
\startdata
NGC 3227      &0.064        &13.2	&1145\\
NGC 3516      &0.048        &18.1	&2624\\
NGC 4151      &1.16         &112.1	&990\\	
NGC 4235      &0.0024       &3.69	&2255\\
NGC 4593      &0.0172       &10.1	&2496\\
NGC 5548      &0.058        &17.7	&5100\\
NGC 6814      &0.0136       &6.25	&1563\\
NGC 7469      &0.058        &48.3	&4807\\
Mrk 6         &0.075        &30.6	&5400\\
Mrk 40        &0.0072       &0.78	&6327\\
Mrk 42        &0.0012       &1.35	&7200\\
Mrk 372       &0.013        &0.69	&9306\\
Mrk 915       &0.046        &4.25	&1769\\
Mrk 1048      &0.021        &12.2	&12934\\
IC 4329A      &0.034        &59.4	&4574\\
\enddata
\tablenotetext{a}{From Whittle 1992}
\tablenotetext{b}{From Quillen et al.~2001}
\tablenotetext{c}{Radial velocity, from SIMBAD}
\end{deluxetable}

\begin{deluxetable}{ccccccc}
\tablewidth{0pc}
\tablecaption{Seyfert 2 Galaxies
\label{tab:Sy2}}
\tablehead{
\colhead{Name}&
\colhead{[\ion{O}{3}] flux\tablenotemark{a}}&
\colhead{$1.6 \micron$ flux\tablenotemark{b}}&
\colhead{$v_{\rm r}$\tablenotemark{c}}&
\colhead{A$_{1.6}$}&
\colhead{$\NH$\tablenotemark{a}}&
\colhead{$A_{1.6}/\NH$}
\\
\colhead{}&
\colhead{$10^{-11} \erg \s^{-1} \cm^{-2}$}&
\colhead{mJy}&
\colhead{$\kms$}&
\colhead{mag}&
\colhead{$10^{20} \cm^{-2}$}&
\colhead{rel.~Gal.}
}
\startdata
NGC 3081	&0.215	&13.4	 &2413	&1.60	 &6600.	   &2.4E-2\\
NGC 4388	&0.374	&0.71	 &2400	&5.25	 &4200.	   &1.2E-1\\
NGC 5128	&0.006	&5.8	 &360	&0.50	 &2000.	   &2.4E-2\\
NGC 5194	&0.228	&0.19	 &463	&7.11	 &7000.	   &9.9E-2\\
NGC 7582	&0.445	&22.6	 &1545	&1.86	 &1240.	   &1.5E-1\\
IC 5063	        &0.353	&0.32	 &3402	&5.89	 &2400.	   &2.4E-1\\
Circinus	&6.970	&4.77	 &439	&6.48	 &4.3E4	   &1.5E-2\\
IRAS 1832-5926	&0.752	&22.7	 &6069	&1.60	 &132.	   &1.2\\
NGC 7319	&0.024	&0.07	 &6611	&4.97	 &3300.	   &1.5E-1\\
NGC 2639	&0.004	&$<0.15$ &3198	&$>3.02$ &4200.	   &$>$ 7.0E-2\\
NGC 4258	&0.262	&$<1.$	 &472\tablenotemark{d}	&$>5.41$	
&1500.	  &$>$ 3.5E-1\\	
NGC 4941	&0.355	&$<0.4$	 &945	&$>6.30$ &4500.	   &$>$ 1.4E-1\\
NGC 1672	&0.077	&$<1.1$	 &1114	&$>3.85$ &63.	   &$>$ 5.9\\
NGC 3079	&0.090	&$<0.1$	 &1125	&$>6.58$ &160.	   &$>$ 4.0\\
NGC 1068	&15.8	&83.6	 &1200	&3.54	 &$>$ 1.E5 &$<$ 3.4E-3\\
NGC 2273	&0.277	&0.32	 &1840	&6.00    &$>$ 1.E5 &$<$ 5.8E-3\\
NGC 5135	&0.614	&0.66	 &3866	&5.50	 &$>$ 1.E4 &$<$ 5.3E-2\\
NGC 5347	&0.100	&0.97	 &2335	&3.83	 &$>$ 1.E4 &$<$ 3.7E-2\\
Mrk 1066	&0.513	&0.51	 &3600	&5.67	 &$>$ 1.E4 &$<$ 5.5E-2\\
NGC 7674	&0.185	&4.39	 &8766	&2.03	 &$>$ 1.E5 &$<$ 2.0E-3\\
Mrk 1210	&0.482	&1.51	 &4049	&4.38	 &$>$ 1.E4 &$<$ 4.2E-2\\
Mrk 477 	&1.238	&0.29	 &11318	&6.43	 &$>$ 1.E4 &$<$ 6.2E-2\\
\enddata
\tablenotetext{a}{From Bassani et al.~1999 and Risaliti et al.~1999} 
\tablenotetext{b}{From Quillen et al.~2001}
\tablenotetext{c}{Radial velocity, from SIMBAD}
\tablenotetext{d}{From Cecil et al.~1992}
\end{deluxetable}

\end{document}